\input{aipcheck}

\documentclass[
    ,final            
  ]
  {aipproc}

\layoutstyle{8x11single}

\usepackage{bm}
\usepackage{amssymb,amsmath}

\begin{document}

\title{Axial Anomaly and Chiral Asymmetry in Magnetized Relativistic Matter}

 
%
%

\author{Igor A. Shovkovy}{
  address={Department of Applied Sciences and Mathematics, Arizona State University, Mesa, Arizona 85212, USA}
}

\begin{abstract}
The generation of a chiral shift parameter in the normal ground state of magnetized relativistic matter 
is discussed. The chiral shift contributes to the axial current density, but does not modify the conventional 
axial anomaly relation. The analysis based on gauge invariant regularization schemes suggests that this 
finding is also valid in gauge theories. We argue that the chiral shift parameter can affect observable  
properties of compact stars. In the regime of heavy ion collisions, the chiral shift parameter can modify 
the key features of the chiral magnetic effect.
\end{abstract}

\maketitle


\section{Introduction}

Many recent theoretical studies \cite{Son-et-al,Metlitski:2005pr,Kharzeev:2007tn,Gorbar:2009bm,Rebhan,
FI1,Basar:2010zd,Kim,Frolov,Gatto:2010qs,Burnier:2011bf} of relativistic matter under extreme conditions, 
e.g., those realized inside compact stars and/or in heavy ion collisions, revealed that matter in a strong 
magnetic field may hold some new surprises. (For lattice studies of relativistic models in strong magnetic 
fields, see Refs.~\cite{Buividovich:2009wi,D'Elia}.) Here we discuss one of them, which was triggered 
by a simingly innocuous finding that a topological contribution to the axial current is already induced in 
the free theory \cite{Metlitski:2005pr}. This lead us to propose that the ground state of such matter is 
characterized by a chiral shift parameter $\Delta$ \cite{Gorbar:2009bm}. The value of $\Delta$ 
determines a relative shift of the longitudinal momenta in the dispersion relations of opposite chirality 
fermions, $k^{3}\to k^{3}\pm\Delta$, where $k^{3}$ is the momentum along the magnetic field. This 
conclusion is approximately valid even in the case of massive particles, provided a sufficiently large 
Fermi surface is formed (i.e., $\mu\gg m$, where $\mu$ is the chemical potential and $m$ is the 
mass of fermions) \cite{Gorbar:2011ya}.

The chiral shift parameter is even under the parity transformation ${\cal P}$ and the charge conjugation 
${\cal C}$, but breaks the time reversal ${\cal T}$ and the rotational symmetry $SO(3)$ down to $SO(2)$. 
Since the global symmetries of dense relativistic matter in an external magnetic field are exactly the 
same, the generation of the chiral shift parameter is expected even in perturbation theory \cite{Gorbar:2009bm}. 

\section{Main results}

{\bf Model.} 
Here we briefly review the dynamics responsible for the generation of the chiral shift 
parameter. While we use a simple Nambu-Jona-Lasinio model, we also envision the generalization 
of the main results to gauge theories. Keeping this in mind, we will utilize a gauge-invariant 
regularization scheme in the analysis below. 

The Lagrangian density of the model reads
\begin{equation}
{\cal L} = \bar\psi \left(iD_\nu+\mu_0\delta_{\nu}^{0}\right)\gamma^\nu \psi
-m_{0}\bar\psi \psi +\frac{G_{\rm int}}{2}\left[\left(\bar\psi \psi\right)^2
+\left(\bar\psi i\gamma^5\psi\right)^2\right],
\label{NJLmodel}
\end{equation}
where $m_{0}$ is the bare fermion mass, $\mu_0$ is the chemical potential, and $G_{\rm int}$ is a 
dimensionful coupling constant. By definition, $\gamma^5\equiv i\gamma^0\gamma^1\gamma^2\gamma^3$. 
The covariant derivative $D_{\nu}=\partial_\nu -i e A_{\nu}$ includes the external gauge field
$A_{\nu}$. 

The structure of the (inverse) fermion propagator is given by the following ansatz:
\begin{equation}
iG^{-1}(u,u^\prime) =\Big[(i\partial_t+\mu)\gamma^0 -
(\bm{\pi}_{\perp}\cdot\bm{\gamma})-\pi^{3}\gamma^3
+ i\tilde{\mu}\gamma^1\gamma^2
+\Delta\gamma^3\gamma^5
-m\Big]\delta^{4}(u- u^\prime),
\label{ginverse}
\end{equation}
where $u=(t,\mathbf{r})$, and the canonical momenta are $\pi_{\perp}^{k} \equiv i \partial^k + e A^k$
(with $k=1,2$) and $\pi^{3} = i \partial^3 =- i \partial_3$. While the spatial components of the gradient 
$\bm{\nabla}$ are given by covariant components $\partial_k$, the spatial components of the vector 
potential $\mathbf{A}$ are identified with the contravariant components $A^k$. We choose the vector 
potential in the Landau gauge, $\mathbf{A}= (0, x B,0)$, where $B$ is the strength of the magnetic field.

In Eq.~(\ref{ginverse}), in addition to the usual tree level terms, two new dynamical 
parameters ($\tilde{\mu}$ and $\Delta$) are included. From the Dirac structure, it should be clear 
that $\tilde{\mu}$ plays the role of an anomalous magnetic moment and $\Delta$ is the chiral 
shift parameter. In the mean-field approximation \cite{Gorbar:2011ya}, there are no solutions 
with a nontrivial $\tilde{\mu}$. So, we take $\tilde{\mu}\equiv 0$ below. 

{\bf Gap equation.} In the mean-field approximation, the gap equation is equivalent to 
the following three equations:
\begin{eqnarray}
\mu =\mu_0 -\frac{1}{2}G_{\rm int} \langle j^{0}\rangle  ,
\qquad
m = m_0 - G_{\rm int}  \langle \bar{\psi}\psi\rangle     ,
\qquad
\Delta = -\frac{1}{2}G_{\rm int}  \langle j^{3}_5\rangle   .
\label{gap-Delta-text} 
\end{eqnarray}
which are solved to determine the three dynamical parameters $\mu$, $m$, and $\Delta$.
Here we will not discuss the vacuum solution, realized at small values of the chemical potential 
($\mu_0\lesssim m_{\rm dyn}/\sqrt{2}$) as the result of the magnetic catalysis \cite{MC1}, but 
concentrate exclusively on the normal ground state with $\Delta\neq 0$, which occurs at nonzero 
fermion density. 

Let us start by analyzing the equation for $\Delta$ in perturbation theory. In the zero order 
approximation, $\mu=\mu_0$ and $\Delta=0$, while the fermion number density $\langle j^0\rangle$ 
and the axial current density $\langle j_5^3\rangle$ are nonzero. In particular, as discussed in 
Ref.~\cite{Metlitski:2005pr}, $\langle j^3_5\rangle_0 =-eB\mu_0/(2\pi^2)$.
(Our convention is such that the electric charge of the electron is $-e$ where $e>0$.) 
To the leading order in the coupling constant, one finds from Eq.~(\ref{gap-Delta-text}) that 
$\Delta \propto G_{\rm int} \langle j^3_5\rangle_0 \neq 0$ and  $\mu - \mu_0 \propto 
G_{\rm int}\langle j^0\rangle_0 \neq 0$. The latter implies that $\mu$ and $\mu_0$ are 
nonequal (in the model at hand, this is a consequence of the Hartree contribution to 
the gap equation). More importantly, we find that a nonzero $\Delta$ is induced. The same 
conclusion is also reached in a more careful analysis of the gap equations, utilizing a 
proper-time regularization \cite{Gorbar:2011ya}. This finding has interesting implications 
for theory and applications. 

{\bf Axial current density.}
As pointed out in Refs.~\cite{Son-et-al,Metlitski:2005pr}, the structure of the topological axial current, 
induced by the lowest Landau level (LLL), is intimately connected with the axial anomaly \cite{ABJ}. 
Then, the important question is whether the form of the induced axial current $\langle j^{3}_{5}\rangle$  
coincides with the result in the theory of noninteracting fermions 
\cite{Son-et-al,Metlitski:2005pr,Rebhan,Hong:2010hi}, or whether it is affected by interactions.

We find that the dynamical generation of the chiral shift parameter $\Delta$ does modify the 
ground state expectation value of the axial current density  \cite{Gorbar:2009bm,Gorbar:2011ya,Gorbar:2010}. 
The corresponding correction to 
the current density was calculated using several different regularization schemes
(including the gauge invariant proper-time and point-splitting ones \cite{Gorbar:2011ya,
Gorbar:2010}). It reads $\langle j_5^3\rangle -\langle j_5^3\rangle_0 \propto a\Lambda^2 \Delta $, 
where $\Lambda$ is a cut-off parameter and $a$ is a dimensionless constant of order $1$.
Formally, this contribution to the the axial current appears to be quadratically divergent 
when $\Lambda\to \infty$. However, it is finite because the solution for $\Delta$ itself
is inversely proportional to $\Lambda^2$. After taking this into account, one finds that the axial 
current density is finite in the continuum limit. The same is expected in renormalizable gauge 
theories, in which $\Delta$ will be a running parameter that falls off quickly enough in 
ultraviolet to render a finite (or, perhaps, even vanishing) correction to the axial current.

{\bf Axial anomaly relation.}
The above result for the axial current density, which gets a correction due to the chiral shift parameter, 
brings up the question whether the conventional axial anomaly relation \cite{ABJ} is affected in any way. 
This issue was studied in Ref.~\cite{Gorbar:2010}, using a gauge invariant point-splitting 
regularization scheme, and it was found that the chiral shift parameter does not modify the 
axial anomaly. This is in agreement with the findings of Refs.~\cite{Son-et-al,Metlitski:2005pr}

 

\section{Discussion}

{\bf Compact stars.}
The chiral shift parameter $\Delta$ may have interesting implications for a degenerate relativistic 
matter in compact stars with a strong magnetic field. As briefly mentioned in the Introduction, the 
states at the Fermi surface can be characterized by their chiralities even for massive fermions if 
$\mu\gg m$. Such states of opposite 
chiralities have different dispersion relations because of a nonzero $\Delta$ and, in effect, form
two different Fermi surfaces, which are asymmetric with respect to the direction along the magnetic 
field \cite{Gorbar:2009bm,Gorbar:2011ya}. It is curious that the LLL and the higher Landau levels 
give opposite contributions to the overall asymmetry of each Fermi surface. For example, the 
left-handed electrons in the LLL occupy only the states with {\em positive} longitudinal momenta 
(pointing in the magnetic field direction). In the higher Landau levels, while the left-handed electrons 
can have both positive and negative longitudinal momenta (as well as both spin projections), there 
are more states with {\em negative} momenta occupied. When $\mu\gg \sqrt{|eB|}$, the relative 
contribution of the LLL to the whole Fermi surface is small, and the overall asymmetry is dominated 
by higher Landau levels. In the opposite regime of superstrong magnetic field, only the LLL is 
occupied and, therefore, the overall asymmetry of the Fermi surface will be reversed. In the intermediate 
regime of a few Landau levels occupied, one should expect a crossover from one regime to the other, 
where the asymmetry goes through zero. 

Taking into account that only left-handed fermions participate in weak interactions, it is natural 
to suggest that neutrinos will scatter asymmetrically off the matter with an asymmetric Fermi 
surface \cite{Gorbar:2009bm}. This provides a possible new mechanism for the pulsar kicks \cite{pulsarkicks}. 
Notably, the mechanism driven by the chiral shift should be quite robust because the neutrino asymmetry 
is built up, rather than washed out  \cite{Kusenko,SagertSchaffner}, by interacting with such matter. 

{\bf Heavy Ion Collisions.}
The chiral shift parameter can also play a role in the regime of heavy ion collisions, where sufficiently 
strong magnetic fields may exist \cite{Skokov:2009qp}. In this case, the value of $\Delta$ is nonzero, 
provided $\mu$ is small but nonvanishing \cite{Gorbar:2011ya}, and 
leads to a dynamical correction to the axial current. Unlike the topological term, the dynamical 
one contains an extra factor of the coupling constant. Therefore, if the qualitative picture 
remains the same at strong coupling in QCD, the effect of the chiral shift parameter on the 
axial current can be substantial. The axial current, in turn, should lead to a modified version of 
the chiral magnetic effect \cite{Gorbar:2011ya}, which does not rely on the initial topological 
charge fluctuations \cite{Kharzeev:2007tn}. In fact, it is the axial current that generates an excess 
of opposite chiral charges around the polar regions of the fireball. These charges trigger two 
``usual" chiral magnetic effects with opposite directions of the vector currents at the opposite 
poles. The inward flows of these electric currents will diffuse inside the fireball, while the outward 
flows will lead to a distinct observational signal: an excess of same sign charges going 
back-to-back. 

\begin{theacknowledgments}
I would like to thank my collaborators Eduard Gorbar and Vladimir Miransky for a fruitful collaboration 
on the topics discussed in this presentation. Also, I would like to thank the organizers of PANIC11 for a 
very well organized conference. This work was supported in part by the U.S. National Science 
Foundation under Grant No. PHY-0969844.
\end{theacknowledgments}

\bibliographystyle{aipproc}   

\end{document}